\documentclass{PoS}
\usepackage{latexsym, amsmath, amssymb}
\DeclareMathOperator{\one}{1\hspace{-2.6pt}\mathsf{I}} 
\newcommand{\txt}[1]{\mathrm{#1}} 
\DeclareMathOperator{\MeeI}{M_\txt{ee}^{-1}}
\DeclareMathOperator{\MooI}{M_\txt{oo}^{-1}}
\DeclareMathOperator{\Mee}{M_\txt{ee}}
\DeclareMathOperator{\Moo}{M_\txt{oo}}
\DeclareMathOperator{\Meo}{M_\txt{eo}}
\DeclareMathOperator{\Moe}{M_\txt{oe}}
\DeclareMathOperator{\Tr}{Tr\,}            

\providecommand{\vary}[1]{\delta\{ #1 \}}   

\title{Non-Hermitian Polynomial Hybrid Monte Carlo }

\ShortTitle{Non-Hermitian Polynomial Hybrid Monte Carlo }

\author{\speaker{Oliver Witzel}%
\\        Humboldt Universit\"at zu Berlin, Institut f\"ur Physik, Newtonstr. 15, 12489 Berlin, Germany\\
        E-mail: \email{witzel@physik.hu-berlin.de}}

\abstract{We report on a new variant of the hybrid Monte Carlo algorithm 
employing a polynomial approximation of the inverse of the non-Hermitian
Dirac-Wilson operator. Our approximation relies on simple and 
stable recurrence relations of complex Chebyshev polynomials. 
First performance figures are presented.}

\FullConference{The XXVI International Symposium on Lattice Field Theory \\
                 July 14 - 19, 2008\\
                 Williamsburg, Virginia, USA}

\begin{document}

\section{Introduction}
Despite the steady progress in developing new machines providing more and more computational resources, two-flavor dynamical fermion simulations remain challenging.  Moving on to large volume simulations \cite{DellaMorte:2007sb} we experienced frequent occurrences of large energy violations within the hybrid Monte Carlo (HMC) update \cite{Duane:1987de} indicating possible algorithmic instabilities and raising worries about reversibility violations.  In \cite{SmallEV} it is pointed out that such instabilities are caused by tiny eigenvalues of the Dirac-Wilson operator. Therefore we propose a new HMC variant reviving the idea to reweight observables and approximate the inverse, non-Hermitian Dirac-Wilson operator by Chebyshev polynomials.  Our new variant is demonstrated using Wilson's lattice action \cite{Wilson:1974sk}
\begin{align}
S(U_\mu, \bar \psi, \psi) = \frac{1}{g_0}\sum_P \Tr\{1-U_P\} + \sum_{x,y}\bar \psi(x) M_{xy} \psi(y), \label{Action}
\end{align}
where the first sum runs over all plaquettes $U_P(x) = U_\mu(x) \cdot U_\nu(x+\hat\mu) \cdot U_\mu(x+\hat \nu)^\dagger\cdot U_\nu(x)^\dagger$ of the gauge field $U_\mu(x)$ and the second sum over all lattice sites. $g_0$ is the strong coupling constant and $M$ is the non-Hermitian Dirac-Wilson operator given in matrix notation.  After integrating out the quark fields (anti-commuting Grassmann variables) $\bar\psi$ and $\psi$ we arrive for the second summand at the fermion determinant $\det\{M\}$. Due to the sparse structure of $M$ it is suitable for even-odd preconditioning, which leads to a factorization of $\det\{M\}$ \cite{Jansen:1996yt}
\begin{align}
\det \{M\} &= \det\{\Mee\} \cdot \det\{\hat M^A\} \\
&= \det\{\Mee\} \cdot \det\{\hat M^S\} \cdot \det\{\Moo\}.
\end{align}
The two possibilities are commonly named \emph{asymmetric}\/ and \emph{symmetric} with the preconditioned operator $\hat M^A = \Moo - \Moe\MeeI \Meo$ and $\hat M^S = \one - \MooI\Moe\MeeI \Meo$, respectively.

Setting up an algorithm to simulate two flavor QCD, we consider the determinant of $M M^\dagger$.  The algorithmic concept of our new HMC variant follows the concept of the polynomial hybrid Monte Carlo (PHMC) by Frezotti and Jansen \cite{Frezzotti:1997ym}. They approximate the Hermitian operator by a root factorization \cite{deForcrand:1996ck} and allow to compensate by a reweighting factor for a possible deviation from importance sampling.  Here we make use of the non-Hermitian, even-odd preconditioned operator  $\hat M = \hat M^S$ or $\hat M^A$, introduce polynomials $P_n \approx \hat M^{-1}$ and incorporate as well a reweighting factor
\begin{align}
\det\{ \hat M \hat M^\dagger \} = \det\{[\hat M P_n] [\hat M P_n]^\dagger\} \cdot [\det\{P_nP_n^\dagger\}]^{-1}. \label{detRW}
\end{align}

In the next section we first motivate our choice of the non-Hermitian Dirac-Wilson operator and show some of its properties. Our approximation of the inverse Dirac-Wilson operator in terms of Chebyshev polynomials is presented in Section \ref{SecApprox} 
and in Section \ref{SecHMC} we introduce the basic steps of our Non-Hermitian Polynomial Hybrid Monte Carlo (NPHMC). The dependence on the polynomial parameters is analyzed in Section \ref{SecPar}.

\section{Non-Hermitian Dirac-Wilson Operator}
Using matrix notation and the hopping parameter representation we can write the Dirac-Wilson operator as $M_{xy} = \delta_{xy} - K_{xy}$, where all interactions are contained in 
\begin{align}
K_{xy} &= \kappa\left( H_{xy} - \frac{i}{2}c_\txt{sw}\sigma_{\mu\nu} {\cal F}_{\mu\nu}\delta_{xy}\right).
\end{align} 
$K_{xy}$ is thus built up by the hopping operator $H_{xy}$ and the $O(a)$ improvement, the Sheikholeslami-Wohlert-term, and is proportional to the hopping parameter $\kappa$.  In \cite{Takeda:2007xu} we present spectral properties of this operator in a set-up with Schr\"odinger functional (SF) boundary conditions (BC).  For our further considerations the following properties are important:
\begin{itemize} 
\item $M$ has a complex spectrum lying in the positive half-plane bounded by an ellipse 
\vspace{-8pt} \item $M^{-1}$ can be approximated recursively by simple and stable polynomials 
\vspace{-8pt} \item We expect a ``good'' approximation with a lower degree polynomial than if approximating the Hermitian operator, which is indicated by a lower condition number\footnote{We thank Tony Kennedy for pointing that out.}
\vspace{-8pt} \item These properties carry over to the even-odd preconditioned operator $\hat M$
\end{itemize}

The latter manifests itself in an exact relation between the eigenvalues $\hat \lambda$ of the preconditioned operator $\hat K$ and the eigenvalues $\lambda$ of $K$ if the $O(a)$ improvement is switched off ($c_\txt{sw} = 0)$
\begin{align}
\hat \lambda(\hat K) = \lambda^2(K). \label{MappingRel}
\end{align}
Assuming next the spectrum of $K$ and $\hat K$, respectively, to fill perfectly an ellipse we parameterize the eigenvalues of $K$ by $\lambda = e \cosh(\vartheta + i \phi)$ with eccentricity $e$ and ``angles'' $\vartheta$ and $\phi$.  From (\ref{MappingRel}) follows now for the spectrum of the preconditioned operator
\begin{align}
\hat e = \hat \delta = e^2/2,
\end{align}
where $\hat e$ is the eccentricity of the ellipse bounding the spectrum of $\hat K$ and $\hat \delta$ marks a positive shift along the real axis. With $O(a)$ improvement turned on, (\ref{MappingRel}) as well as the drawn conclusions hold only approximately. 

\begin{figure}[ht]
\centering
\includegraphics[width=0.95\textwidth,clip]{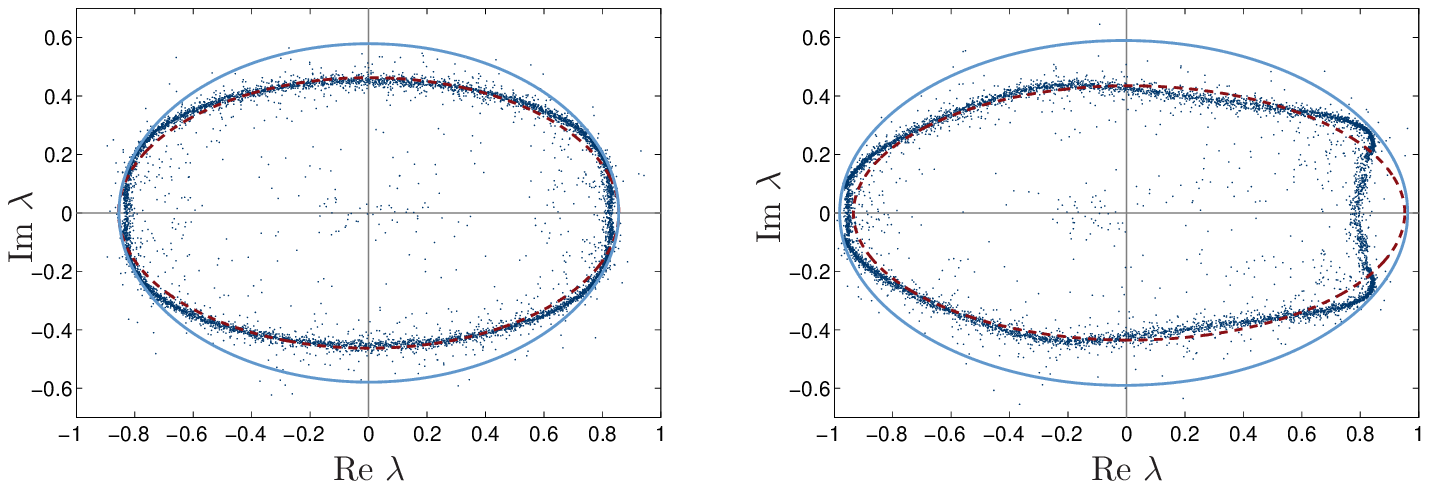}
\includegraphics[width=0.95\textwidth,clip]{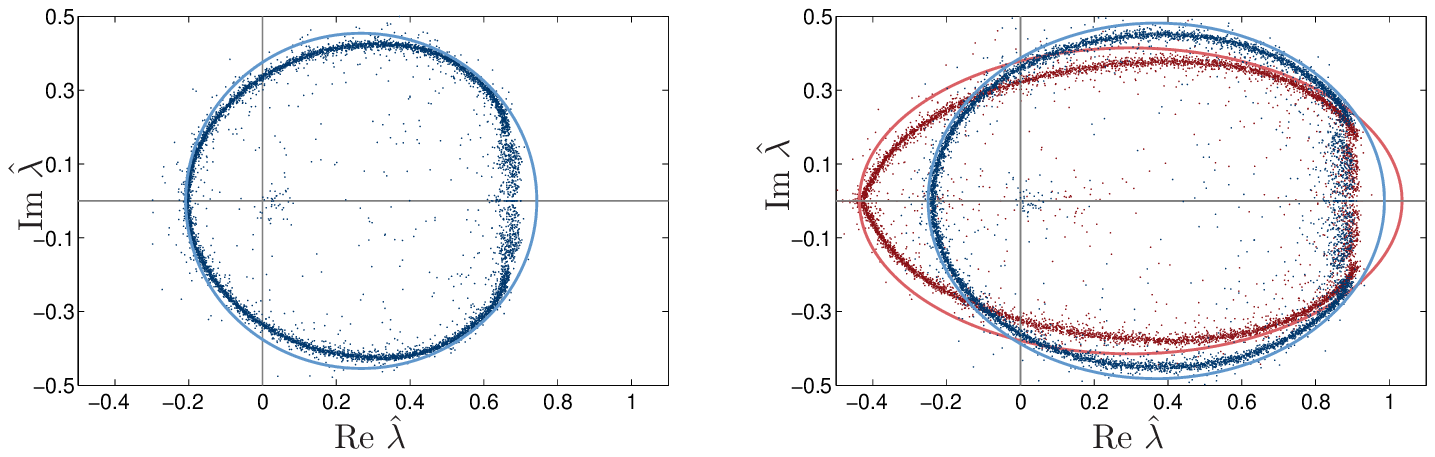}
\caption{Computing the spectral boundary of $\one - M$ (upper plots) and $\one - \hat M$ (lower plots) by the Lanczos method on a set of 50 pure-gauge configurations: $8^4$ lattice at $\beta = 6.0$ and $\kappa = 0.13458$ in the  SF. Left without / right with clover-term; lower right plot shows in red asymmetric and in blue symmetric even-odd preconditioning.}\label{Spectra}
\end{figure}

Computing the spectral boundary numerically by using the complex Lanczos method  we can nicely visualize the shift and the advantage due to even-odd preconditioning (see Fig.~\ref{Spectra}).  Moreover we see that with Sheikholeslami-Wohlert term the symmetric version of even-odd preconditioning is superior because it leads to a more compact and round spectrum. In addition one can verify certain symmetries of the operator (see e.g.~\cite{Gattringer:1997ci}): without the clover term and if all dimensions of the lattice are even, the spectrum exhibits a mirror symmetry under sign flip and has complex pairs of eigenvalues ($\gamma_5$ Hermiticity). With clover term the first is not present, but we still find complex pairs of eigenvalues.

\section{Chebyshev Approximation}\label{SecApprox}
In order to find a polynomial approximation of the inverse, non-Hermitian Dirac-Wilson operator $\hat M$ we first introduce a ``small quantity'' (remainder)
\begin{align}
R_{n+1}(\hat M) = \one - \hat M P_n. \label{Remainder}
\end{align}
By construction $R_{n+1}$ is small on an elliptical region of the spectrum of $\hat M$ and suitable for an approximation by scaled and translated Chebyshev polynomials as introduced by Manteuffel \cite{Manteuffel:1977a}
\begin{align}
R_{n+1}(\hat M) = \frac{T_{n+1}(\hat K/\hat e) }{T_{n+1}(\hat d/\hat e)}. \label{Cheby}
\end{align}
Here $\hat d = 1 + \hat \delta$ is ideally such that the spectrum of $\hat K$ is origin centered and hence $\hat M = \hat d - \hat K$.  The spectral region approximated is bounded by an ellipse with eccentricity $\hat e$.   Eq.~(\ref{Cheby}) together with the well-known recurrence relation for Chebyshev polynomials,
\begin{align}
T_{n+1}(z) = 2 z T_n(z) - T_{n-1}(z)\quad\txt{with}\quad T_1=z\quad \txt{and}\quad T_0 =1,
\end{align}
provides the key to obtain first a recursive description for the $R_{n+1}$ and by (\ref{Remainder}) also for our sought polynomials $P_n$ \cite{PolyNotes},
\begin{align} 
R_{n+1} &= a_n \hat K R_n + (1-\hat d a_n) R_{n-1} &\txt{with}\quad& R_1=\hat K/\hat d &\txt{and}\quad& R_0 =\one, \label{Rn+1}\\
P_{n} &= a_n(\one+\hat K P_{n-1}) + (1-\hat d a_n)P_{n-2} &\txt{with}\quad& P_1=a_1(\one+ \hat K /\hat d) &\txt{and}\quad& P_0 =\one/\hat d,\label{Pn}
\end{align}
The real coefficients $a_n$ are given by 
\begin{align}
a_n = (\hat d - a_{n-1}\hat e^2/4)^{-1} \qquad \txt{with}\quad a_1  = \hat d(\hat d^2-\hat e^2/2)^{-1}.
\end{align}
and converge to $\lim_{n \to \infty}\; a_n = 2 (\hat d - \hat e \sqrt{\hat d^2/\hat e^2 -1})/\hat e^2$.  
 Concerning the two recurrence relations (\ref{Rn+1}) and (\ref{Pn}) we like to emphasize: although given as matrix relations, (\ref{Rn+1}) and (\ref{Pn}) lead to repeated, numerically cheap matrix times vector multiplications. Furthermore, the inverting polynomial is obtained from numerically stable and simple two-step recursions.\cite{Witzel:2008phd}

\section{NPHMC update}\label{SecHMC}
To generate configurations with the Boltzmann weight $\exp\{-S\}$ by a NPHMC update, we manipulate the action (\ref{Action}) accordingly 
\begin{align}
S = \frac{1}{g_0}\sum_P \Tr\{1-U_P\}  + 2[\ln \det\{\Mee\} + \ln\det\{\Moo\}] + \underbrace{\phi^\dagger P_n^\dagger(\hat M) P_n(\hat M) \phi}_{S_b}, \label{ActionHMC}
\end{align}
where from now on we choose $\hat M = \hat M^S$ since $\hat M^S$ is superior to $\hat M^A$.  The first term is the unchanged contribution from the gauge fields, it follows the determinant contribution due to even-odd preconditioning and the last term gives rise to the bosonic contribution obtained by estimating the second factor of the right hand side in (\ref{detRW}) as bosonic integral.  We take care of the first factor in (\ref{detRW}), $\det\{[\hat M P_n] [\hat M P_n]^\dagger\}$, by calculating a stochastic estimate using random Gaussian fields $\eta_C$
\begin{align}
C &= \exp\left\{\eta_C^\dagger [ \one  - (P_n^\dagger \hat M^\dagger \hat M P_n)^{-1}] \eta_C\right\}. \label{RWF}
\end{align}
This term is only computed when measuring observables.  The general procedure of an HMC update is unchanged.  Differences occur in the way the bosonic contribution $S_b$ is treated. 

At the beginning of a trajectory the pseudo-fermion fields $\phi$ have to be generated with the correct distribution by inverting $P_n$ on a random Gaussian field $\eta$
\begin{align}
\phi  = P_n^{-1} \eta = (\one - R_{n+1})^{-1}\hat M \eta.
\end{align}
Here it is important to use the second relation and by that invert a well conditioned matrix needing only a little number of expensive iteration steps.  Next we compute the variation $\vary{S_b}$ as it is required to integrate the equations of motions. Varying each occurrence of $\hat K$ and reorganizing the expressions we yield 
\begin{align}
\vary{S_b} &= \sum_{l=1}^n \left[ \xi^\dagger_{2n-l}a_l \vary{\hat K}\chi_{l-1}+\txt{H.c.}\right] &\\
\intertext{with}
\chi_j &= a_j \phi + a_j \hat K \chi_{j-1}+(1-\hat d a_j)\chi_{j-2}; 
&\chi_1 = a_1 (\one +\hat K/\hat d)\phi;& & \chi_0 = \phi/d& \\
\xi^\dagger_{n+j}&= \xi^\dagger_{n+j-1}\hat K a_{n-j+1} + \xi^\dagger_{n+j-2}(1-\hat d a_{n-j+2});
&\xi^\dagger_{n+1} = \chi_n^\dagger \hat K a_n;\qquad& &\xi^\dagger_n = \chi_n^\dagger.\;&
\end{align}
Finally, the reweighting factor $C$ is estimated requiring again the inversion of a well-conditioned matrix. Of course also here we can replace $\hat M P_n$ by $\one - R_{n+1}$. For further details see \cite{Witzel:2008phd}.

\section{Performance Tests} \label{SecPar}
Testing the proposed new HMC variant we investigate the dependence on the three polynomial parameters $\hat \delta$, $\hat e$ and $n$ on an $8^4$ lattice at $\beta=6.0$ and $\kappa = 0.13458$.  The tests are performed by keeping two parameters fixed, while varying the third one and monitoring as observables: the mean value of the correction factor $\langle C \rangle$, the relative width of its distribution $\varsigma_C =  \sqrt{\langle C^2\rangle - \langle C\rangle^2}/ \langle C \rangle$ as well as  the number of conjugate gradient (CG) iterations required to compute the correction factor, $\#(\txt{iterations\; CG})$. 

\begin{figure}[ht]
\centering
\includegraphics[width=0.9\textwidth,clip]{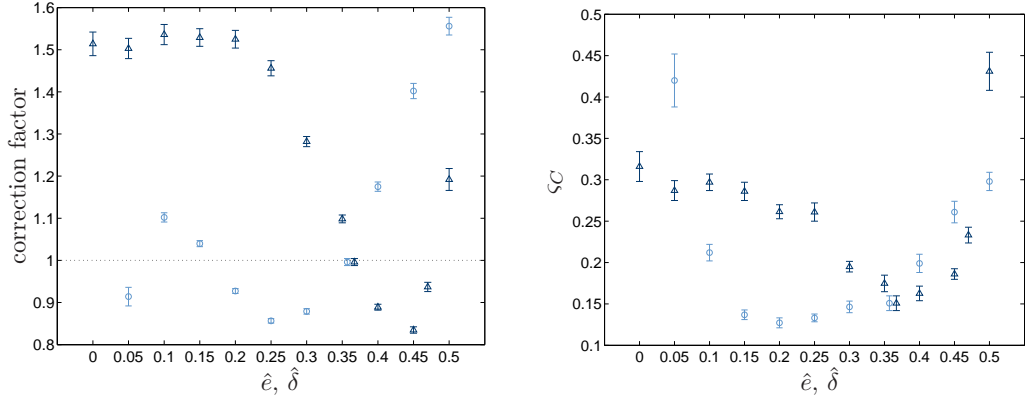}
\caption{Dependence of the correction factor $C$ and the width of its distribution $\varsigma_C$ on $\hat \delta$ (light blue $\circ$) and $\hat e$ (dark blue {\footnotesize $\triangle$}), respectively.}
\label{e_delta1}
\end{figure}

\begin{figure}[hbt]
\centering \vspace{-15pt}
\includegraphics[width=0.4\textwidth,clip]{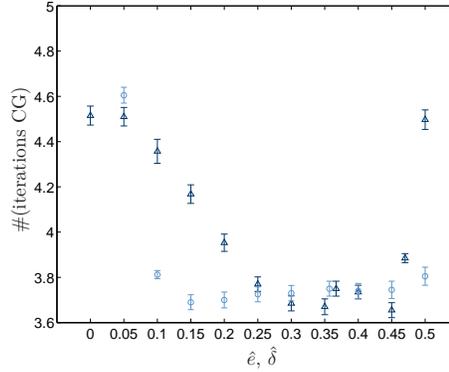}
\caption{Dependence of the number of CF iterations on $\hat \delta$ (light blue $\circ$) and $\hat e$ (dark blue {\footnotesize $\triangle$}), respectively.}
\label{e_delta2}
\end{figure}

As can be seen by looking at Figs.~\ref{e_delta1} and \ref{e_delta2} the two parameters specifying the elliptical region of the approximating polynomial favor $\hat e \approx \hat \delta \approx 0.36$ but do not require a sophisticated fine tuning.  This value is in good agreement with the ones concluded from the bounding ellipse shown in Fig.~\ref{Spectra} in case of symmetric even-odd preconditioning. 

\begin{figure}[ht]
\centering
\includegraphics[width=0.5\textwidth,clip]{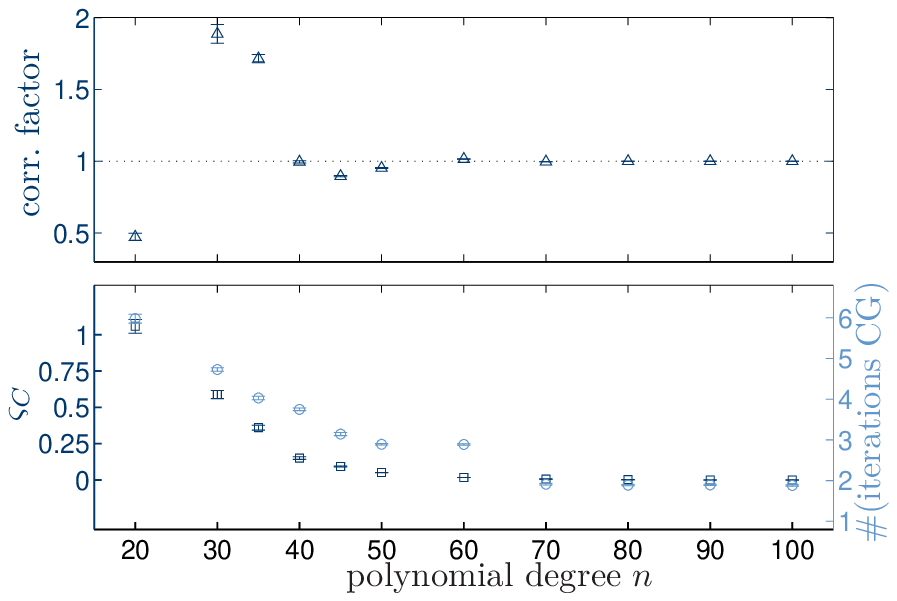}
\caption{Dependence of $C$ ({\footnotesize $\triangle$}), $\varsigma_C$ ($\Box$) and $\#(\txt{iterations\; CG})\; (\circ)$ on the polynomial degree $n$.}
\label{nDep}
\end{figure}

Turning to the dependence on the polynomial degree $n$ (Fig.~\ref{nDep}) we find a much stronger dependence. This parameter is mainly responsible for the quality of our approximation. Hence it affects the cost as well as the noisiness of the correction factor.  For a higher degree polynomial $C$ approaches one, while $\varsigma_C$  and $\#(\txt{iterations\; CG)}$ go to zero. Using the $4 n \cdot \#(\txt{iterations\; CG)}$ as a coarse estimate for the cost, a good choice of the polynomial degree -- for this set-up -- is $n = 50$.

\vspace{-6pt}
\section{Conclusion}\vspace{-5pt}
Employing the scaled and translated Chebyshev polynomials to approximate the inverse, non-Hermitian Dirac-Wilson operators allows to derive a variant of the HMC update algorithm based on simple and stable recurrence relations.  The approximation does not require a fine tuning of the parameters specifying the elliptical approximation region, whereas the polynomial degree $n$ is important for the quality of the approximation and the numerical cost.

A conclusive comparison between the performance of different HMC-type algorithm has not been performed yet. However, the first data indicate that the 1-pseudo-fermion NPHMC is slightly superior than a standard 1-pseudo-fermion HMC, but inferior to an HMC version incorporating the Hasenbusch-trick \cite{Hasenbusch:2001ne} and multiple time scale integration \cite{Urbach:2005ji}.  Constructing a NPHMC with two pseudo-fermions by means of the Hasenbusch-trick is possible.  Unfortunately, this forces an involved tuning of the polynomial degrees and appears to be not too promising. 

As a byproduct of our studies of the non-Hermitian Dirac-Wilson operator, we found that the symmetric even-odd preconditioned operator is advantageous because of a more compact spectrum.

\paragraph{Acknowledgement\\}
The author thanks B.~Bunk and U.~Wolff for fruitful discussions and the APE team at DESY Zeuthen for technical support. This work is supported by the DFG within the SFB/TR 9.

\end{document}